\documentclass[sigconf,nonacm]{acmart}
\AtBeginDocument{%
  }

\setcopyright{acmlicensed}
\copyrightyear{2018}
\acmYear{2018}
\acmDOI{XXXXXXX.XXXXXXX}
\acmConference[Conference acronym 'XX]{Make sure to enter the correct
  conference title from your rights confirmation email}{June 03--05,
  2018}{Woodstock, NY}
\acmISBN{978-1-4503-XXXX-X/2018/06}
\usepackage{hyperref}
\usepackage{multicol}

\usepackage{xspace}
\usepackage{graphicx}
\usepackage[frozencache,cachedir=.]{minted} 
\usepackage{caption}
\usepackage{listings}
\usepackage{array}
\usepackage{hhline}
\usepackage{textcomp}
\definecolor{hl}{HTML}{FFF9C4}
\usepackage{svg}
\usepackage[greek,english]{babel}

\newcommand{\projName}{\textsc{Mnimi}\xspace}

\begin{document}

\title{Statistical Independence Aware Caching for LLM Workflows}

\author{Yihan Dai}
\email{yhdai25@stu.pku.edu.cn}
\affiliation{%
  \institution{Peking University}
  \city{Beijing}
  \country{China}
}

\author{Dimitrios Stamatios Bouras}
\email{2501112125@stu.pku.edu.cn}
\affiliation{%
  \institution{Peking University}
  \city{Beijing}
  \country{China}
}

\author{Haoxiang Jia}
\email{haoxiangjia@stu.pku.edu.cn}
\affiliation{%
  \institution{Peking University}
  \city{Beijing}
  \country{China}
}

\author{Sergey Mechtaev}
\email{mechtaev@pku.edu.cn}
\affiliation{%
  \institution{Peking University}
  \city{Beijing}
  \country{China}}

\renewcommand{\shortauthors}{Dai et al.}

\begin{abstract}
Large language models (LLMs) inference is both expensive and slow. Local caching of responses offers a practical solution to reduce the cost and latency of LLM queries. In research contexts, caching also enhances reproducibility and provides flexibility for experimentation. However, naive reuse of cached responses compromises statistical independence, a critical property for probabilistic workflows. In applications of LLM for code, it underpins performance metrics such as Pass@k and uncertainty estimation, as well as algorithms like program repair loops and retries. Existing LLM caching systems lack ways to enforce statistical independence constraints. To address this, we introduce \projName, a cache design pattern that supports modular LLM workflows while ensuring statistical integrity at the component level. Its core innovation lies in encapsulating statistical constraints within the type of LLM references, allowing users to manage and transform these types according to the scope and requirements of their algorithm. We implemented this design pattern in Python using a combination of decorators and iterators over infinite sequences. A case study on SpecFix, an recent automated program specification repair system, highlights how \projName improves reproducibility, ease of debugging, time and cost efficiency while preserving statistical correctness.
\end{abstract}



\maketitle

\section{Introduction}

\begin{figure*}[htbp]
  \begin{center}
    \includesvg[width=\textwidth]{example}
  \end{center} 
  \caption{An example showing the different behaviors of \texttt{Independent} vs. \texttt{Repeatable}. In the ``Iterator State Transition'' column, the first sub-row of it shows the sequence state returned by calling \texttt{sample} and the second sub-row shows the state after iterating to retrieve elements, while ``\texttt{→}'' denotes the pointer to the next element inside the iterator.}
  \vspace{-3mm}
  \label{fig:example}
\end{figure*}

The use of LLMs in software engineering is rapidly expanding, however this comes with substantial costs in time, money, and energy due to LLM inference. LLM backend optimizations~\cite{kwon2023vllm} alleviate these issues but are model-specific, requiring extensive engineering or proprietary hardware access. A practical alternative is client-side caching of LLM responses~\cite{bang-2023-gptcache, meancache2024, potamitiscache}, which avoids redundant queries, improving speed, cost-efficiency, and enabling deterministic debugging, without altering the LLM. In research contexts, it enhances reproducibility and facilitates experimentation.

A typical client-side caching approach, as implemented in the popular LangChain framework~\cite{langchain2023}, associates each prompt with a single LLM response. Whenever the same prompt is encountered again, the cached response is returned instead of generating a new one. This design conflicts with statistical independence required by many algorithms. In LLMs for code, widely adopted metrics like Pass@k~\cite{chen2021evaluatinglargelanguagemodels} rely on generating statistically independent samples to estimate the probability of a solution's success. Similarly, methods for calibrating model uncertainty~\cite{hobelsberger2025systematicevaluationuncertaintyestimation, lyu2025calibrating, manakul2023selfcheckgptzeroresourceblackboxhallucination}, as well as iterative repair and refinement pipelines~\cite{jia2025automatedrepairambiguousproblem, madaan2023selfrefineiterativerefinementselffeedback, shinn2023reflexionlanguageagentsverbal}, inherently depend on the model's ability to produce varied outputs for the same prompt. Meanwhile, a naive cache collapses this necessary variation.

A straightforward approach to address this problem is to cache a sequence of independent responses for each prompt and replay them during every application run, instead of caching a single response. While this ensures statistical independence within each run, it has practical drawbacks. First, it is inefficient because it does not reuse samples within the run. Second, it undermines reproducibility and debugging, as even small changes to the program can disrupt the sequence of responses being replayed. Cache Saver~\cite{potamitiscache} alleviated this issue by introducing explicit namespaces, where each namespace stores independent responses for a given prompt. However, this approach prohibits reuse within a namespace by design, limiting flexibility for users in controlling granularity of sample reuse. Furthermore, LLM workflows are often modular and deeply nested and therefore sampling constraints are also modular and nested. The requirement to explicitly maintain statistical independence namespaces harms modularity.

To address these limitations, we introduce \projName \footnote{\projName\ derives from the Greek word \textgreek{μνήμη} (\textit{mnēmē}), meaning ``memory.'' In Greek mythology, Mnēmē  is one of the three primordial Muses, symbolizing remembrance, the foundation of knowledge and creativity. This aligns our goal to bring structured memory to LLM workflows, balancing reuse and independence.}, a new caching design pattern that supports modular LLM workflows, while ensuring statistical integrity at the component level. At its core, the idea is to encapsulate statistical independence constraints as types of LLM references: \texttt{Repeatable}, ensuring repeated responses for each LLM query, and \texttt{Independent}, ensuring independent responses for each query. This design pattern promotes modularity by enabling transformations of LLM reference types through decorators~\cite{gamma1995design}, effectively allowing the user to specify statistical constraints as easily as ``wrapping a gift''. These layers form a coherent hierarchy governing stability within a program scope, diversity across scopes, and continuity across runs --- capturing the complete statistical semantics that a typical application requires.

To show \projName's applicability in the context of LLM for code, we conducted a case study on SpecFix~\cite{jia2025automatedrepairambiguousproblem}, a recent tool for repairing ambiguous programming problem descriptions. Its heavy reliance on stochastic LLM calls makes runs expensive and hard-to-replicate, with variations in generated tests, programs, and repaired description. A naive LLM caching cannot be adopted, because of SpecFix's reliance on semantic entropy~\cite{kuhn2023semantic} and Pass@k computation, as well as its main repair loop, all requiring independent samples. To resolve this tension, we replaced the original LLM calls in SpecFix with \projName, while carefully encoding the statistical independence constraints via the type of references to the LLM object. Experiments show that \projName improves execution time and inference cost over the original tool, and enables bit-reproducibility, while requiring only minor changes to SpecFix source code.

A Python implementation of \projName, that supports multiple providers, cache slicing (extracting a subset of cache), and batch sampling, as well as our SpecFix integration are available at
\begin{center}
  \url{https://github.com/msv-lab/mnimi}
\end{center}

\section{\projName}

The challenge with statistical independence arises when an application repeatedly queries an LLM using the same prompt, i.e. samples multiple values from the same distribution. The following Python snippet illustrates an LLM-based guessing game
\begin{minted}[breaklines, escapeinside=||, frame=single, fontsize=\small]{python}
template = "Choose an integer from {interval} for {user} to guess."
user = ['Alice', 'Bob', 'Alice']
interval = "[1, 1000]"
responses = []
for user in users:
  prompt = template.format(user=user, interval=interval)
  response = model.sample(prompt) # queries an LLM
  responses.append(response)
\end{minted}
In this example, a shared prompt template is used across different games to generate guessing numbers within the given interval. A key characteristic of the input \texttt{users} is that it may include duplicate usernames (e.g.,  \texttt{\textquotesingle Alice\textquotesingle}), reflecting the scenario where a single user engages with multiple distinct games.

The critical issue lies in whether the responses for duplicated usernames (e.g., Alice's first and third entries) are repeated or independent, as this directly impacts Alice's strategy for guessing the number. Assume that a fair game requires that responses are independent, that is, every time an integer is requested from the LLM for a specific user, it will ensure that the response for this time is independent of all previous responses\footnote{note that independence does not imply uniqueness because of stochasticity}. However, if the application is changed to use a naive cache that simply returns a cached response to the same prompt, the responses will then be repeated. Since statistical independence is a prerequisite of the program above, using naive cache contradicts the intended semantics. Thus, the use of a cache in this scenario requires encoding and respecting algorithm's sample independence constraints.

Let's extend the example so that each user play several rounds of the game, possibly guessing numbers from different intervals:

\begin{minted}[breaklines, escapeinside=||, frame=single, fontsize=\small]{python}
ints = ["[1, 1000]", "[1001, 2000]", "[1, 1000]"]
for user in users:
  for interval in ints:
    prompt = template.format(user=user,interval=interval)
    response = model.sample(prompt)
    responses.append(response)
\end{minted}

\begin{figure*}[htbp]
  \begin{center}
    \includesvg[width=0.75\textwidth]{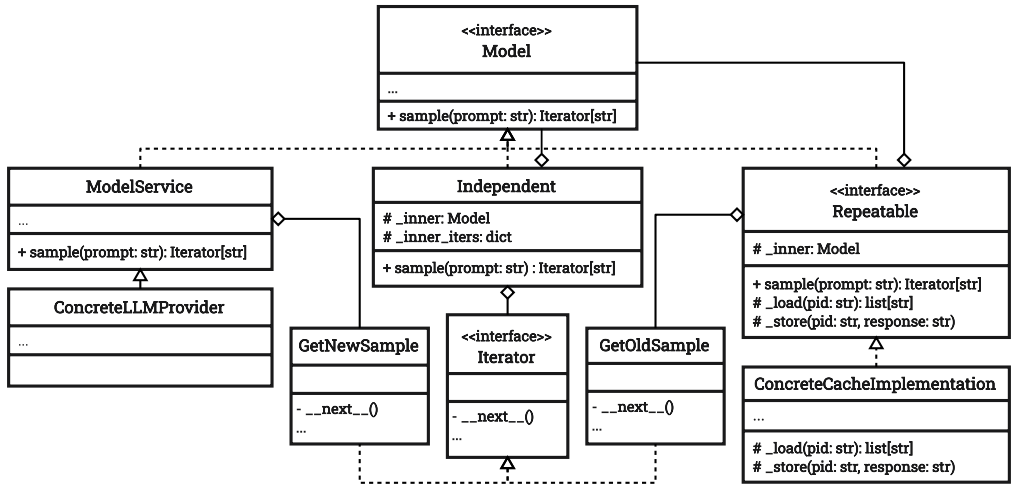}
  \end{center} 
  \caption{Structure of \projName design pattern that combines decorators and iterators over infinite sequences of responses to encode statistical independence constraints in a modular fashion. \projName provides two concrete implementations of \texttt{Repeatable}: an in-memory cache \texttt{InMemory} and an on-disk cache \texttt{Persistent}.}
  \vspace{-2mm}
  \label{fig:structure}
\end{figure*}

Assume we ensured independence across the outer loop iterations; a new question arises: within a single iteration of the outer loop, should the responses across rounds be independent for the same interval, e.g. for the repeated interval \texttt{[1,1000]}? Let the game dictate that across the rounds, users must always guess the same number if the interval remains unchanged. In other words, sampling within each iteration should be \emph{repeatable}. To achieve this, an ad-hoc caching mechanism could be implemented, such as maintaining a dictionary to store the samples corresponding to each element in \texttt{ints}, reusing them when they appear a second time. However, this approach compromises readability and modularity, as it tightly couples the program logic to the cache implementation. Additionally, if different rounds require different numbers of samples, the complexity would increase further, potentially introducing bugs. Thus, to ensure consistent use of cached samples, one must specify scopes of the algorithm in which samples are expected to repeat, or be independent.



\subsection{Cache Design Pattern}

At \projName's core is the \texttt{Model} interface providing a single method
\begin{minted}[breaklines, escapeinside=||, fontsize=\small]{python}
def sample(self, prompt: str) -> Iterator[str]
\end{minted}
which takes in a prompt and returns an infinite sequence of i.i.d. responses from an LLM. The independence of samples within the sequence must be guaranteed by all implementation of this interface. In particular, we introduce three implementations of \texttt{Model} shown in Figure~\ref{fig:structure}. Firstly, \texttt{ModelService} returns an iterator that queries the model provider for new samples, without caching. In this context, sample independence must be ensured by the provider.

Secondly, to realize a caching mechanism, we introduce a decorator \texttt{Repeatable}. It wraps an existing model so that its \texttt{sample} method produces a specialized type of iterator, \texttt{GetOldSample}. This iterator retrieves responses from a cache associated with each prompt; in case of cache misses, samples are obtained from the inner model, and added to the cache. Importantly, \texttt{Repeatable} still ensures independence of responses in the lazy sequence from a single call of \texttt{sample}. Instead, the repeatability is guaranteed across different calls of \texttt{sample}. To achieve that, \texttt{Repeatable} returns a new iterator over the same sequence for each call of \texttt{sample} with the same prompt. Figure~\ref{fig:example} illustrates the impact of repeatability (Line 5), where even though \texttt{n3} has consumed \texttt{\textquotedbl 131\textquotedbl}, the iterator returned to \texttt{n4\_list} still begins at \texttt{\textquotedbl 131\textquotedbl}. While \texttt{Repeatable} serves as an implementation of \texttt{Model}, it also abstracts the storage layer through the \texttt{store} and \texttt{load} methods, which must be implemented by concrete cache backends in \texttt{ConcreteCacheImplementation}.

Finally, to achieve statistical independence, we introduce a decorator \texttt{Independent}. In contrast to \texttt{Repeatable}, it ensures that each call to \texttt{sample} produce an independent sequence of responses. To achieve that, the implementation of \texttt{sample} for \texttt{Independent} merely always returns the same iterator for the same prompt. Since the iterator is shared, it is consumptive --- responses used by one call will not be reused by others, making the effect stateful. Importantly, \texttt{Independent} itself neither generates nor stores responses; rather, it only defines a strategy for consuming them from the inner model. The use of \texttt{Independent} in Figure~\ref{fig:example} in Line 8 ensures that since \texttt{n5\_list} has already used \texttt{\textquotedbl 131\textquotedbl} and \texttt{\textquotedbl 561\textquotedbl}, the iterator for \texttt{n6\_list} skips them and resamples, yielding \texttt{\textquotedbl 452\textquotedbl} and \texttt{\textquotedbl 980\textquotedbl}.

\subsection{Design Pattern Application Examples}
\label{sec:examples}

Here, we extend the example from the beginning of this section to three cases, illustrating how our design pattern elegantly handles successive changes in sampling constraints. At first, we assume that we have  already cached a sequence of independent responses for each prompt (showing prompt template parameters only):

\begin{minted}[breaklines, escapeinside=||, fontsize=\small]{python}
template("Alice", "[1, 1000]"):     ["2", "68", "109", "12"] 
template("Bob",   "[1, 1000]"):     ["297", "573"]
template("Alice", "[1001, 2000]"):  ["1393", "1002"] 
template("Bob",   "[1001, 2000]"):  ["1740"]
\end{minted}

First, assume the user must be given the same number in every round, even in different games:

\begin{minted}[breaklines, escapeinside=||, frame=single, fontsize=\small]{python}
model = Persistent(model, "/cache/path")
responses = []
for user in users:
  for interval in ints:
    prompt = template.format(user=user,interval=interval)
    response = next(model.sample(prompt))
    responses.append(response)
assert responses == ["2", "1393", "2", "297", "1740", "297", "2", "1393", "2"]
\end{minted}

\noindent where \texttt{Persistent} is an subclass of \texttt{Repeatable} that caches responses on-disk. Now, assume a user is given an independent number every time they request one:

\begin{minted}[breaklines, escapeinside=||, frame=single, fontsize=\small]{python}
model = Persistent(model, "/cache/path")  
model = Independent(model)
responses = []
for user in users:
  for interval in ints:
    prompt = template.format(user=user,interval=interval)
    response = next(model.sample(prompt))
    responses.append(response)
assert responses == ["2", "1393", "68", "297", "1740", "573", "109", "1002", "12"]
\end{minted}

While the two examples above show trivial caching semantics supported by previous systems, \projName provides more fine-grained control over sample independence. To show it, assume a user must be given the same number in every round for the same interval, but across different games, corresponding to different iterations of the outer loop, numbers given to the same user must be independent, it requires composing repeatable and independent layers:

\begin{minted}[breaklines, escapeinside=||, frame=single, fontsize=\small]{python}
model = Persistent(model, "/cache/path")  
model = Independent(model)
responses = []
for user in users:
  rep_model = InMemory(model)
  for interval in ints:
    prompt = template.format(user=user,interval=interval)
    response = next(rep_model.sample(prompt))
    responses.append(response)
assert responses == ["2", "1393", "2", "297", "1740", "297", "68", "1002", "68"]
\end{minted}

\noindent where \texttt{InMemory}, a subclass of \texttt{Repeatable} with in-memory cache, ensures deterministic reuse within the loop, \texttt{Independent} ensures independent sampling across loop iterations, and \texttt{Persistent} provides on-disk cache for reproducibility across runs. Here, statistical independence constraints are elegantly encoded through LLM reference types. In contrast, systems like Cache\ Saver that reply on explicit namespaces prohibit intra-namespace reuse of samples, which prevents expressing repeatability within each loop iteration.

\section{Case Study: \projName Integration in SpecFix}

\begin{table*}[ht]
\centering
\caption{
Token usage, runtime, and cost of LLM queries in SpecFix with and without \projName caching (100 HumanEval+ tasks, \texttt{gpt-4.1-mini}, temperature~1.0). C is the number of generated programs, E is the number of generated test. 
\underline{Baseline} does not read from the cache but records all responses for future reuse, with its results reflecting how Specfix behaved originally.
\underline{No cache} runs ignore caching entirely (no reads or writes).
\underline{First cached} reuses stored results from the baseline, paying only for unseen generations,
and \underline{Cached replay} deterministically replays all queries at zero cost.
}

\renewcommand{\arraystretch}{1.2}
\begin{tabular}{|l|r|r|r|r|r|}
\hline
\textbf{Run configuration} & \textbf{Prompt tokens} & \textbf{Completion tokens} & \textbf{Total tokens} & \textbf{API time (s)} & \textbf{Total cost (\$)} \\ \hline
Baseline ($c{=}20$, $e{=}10$)      & 97{,}669  & 319{,}814  & 417{,}483  & 1{,}716.1  & 1.10 \\ \hline
Cached replay ($c{=}20$, $e{=}10$)           & \textbf{0}  & \textbf{0}  & \textbf{0}  & \textbf{0.0}  & \textbf{0.00} \\ \hline
No cache ($c{=}40$, $e{=}10$)                & 100{,}733  & 500{,}792  & 601{,}525  & 2{,}344.9  & 1.68 \\ \hline
First cached ($c{=}40$, $e{=}10$)            & 50{,}885  & 250{,}349  & 301{,}234  & 867.5  & 0.84 \\ \hline
Cached replay ($c{=}40$, $e{=}10$)           & \textbf{0}  & \textbf{0}  & \textbf{0}  & \textbf{0.0}  & \textbf{0.00} \\ \hline
No cache ($c{=}20$, $e{=}20$)                & 94{,}207  & 402{,}743  & 496{,}950  & 2{,}234.6  & 1.36 \\ \hline
First cached ($c{=}20$, $e{=}20$)            & 2{,}941  & 32{,}872  & 35{,}813  & 93.9  & 0.11 \\ \hline
Cached replay ($c{=}20$, $e{=}20$)           & \textbf{0}  & \textbf{0}  & \textbf{0}  & \textbf{0.0}  & \textbf{0.00} \\ \hline
\end{tabular}
\label{tab:mnimi_experiment}
\end{table*}
SpecFix~\cite{jia2025automatedrepairambiguousproblem} automatically repairs ambiguous programming problem descriptions to improve LLM-based code generation. It samples multiple candidate programs based on a given natural-language description, clusters their behaviors, and evaluates interpretive ambiguity using metrics such as semantic entropy and example consistency. The pipeline consists of several stages: test generation, program sampling, and iterative repair, all of which rely heavily on a large number of stochastic LLM calls. As a result, the process can produce varying outputs --- including tests, programs, and final repaired descriptions --- across different runs, making it expensive to reproduce and hard to debug. Meanwhile, the metrics driving SpecFix, such as semantic entropy and Pass@K, are fundamentally statistical and depend on independent samples. Furthermore, SpecFix's repair loop necessitates generation of independent repairs. These aspects make naive caching not applicable for SpecFix.

We applied \projName to SpecFix's source code, systematically encoding the required statistical independence constraints using LLM reference types. Each caching layer in SpecFix serves a distinct semantic purpose tailored to the statistical requirements of the repair process. Listing~\ref{code_specfix} presents a simplified version of SpecFix's repair loop, which iteratively refines a given description over a maximum of \texttt{max\_attempts}. If the score for a repaired description fails to improve, the repair is discarded. This last aspect of the algorithm introduces specific requirements for statistical independence in LLM samples. While stable score computation for each description is desirable --- achieved through the \texttt{Persistent} caching layer, which consistently provides the same sets of programs and tests for the same description --- each repair attempt must remain independent. This independence is crucial to prevent the algorithm from becoming stuck in repeated unsuccessful attempts when the score fails to improve. Thus, we utilize an \texttt{Independent} model reference, which guarantees that each repair attempt explores fresh samples.

\begin{listing}[H] 
\caption{Statistical independence constraints in SpecFix repair loop encoded via \projName decorators.} 
\label{code_specfix}
\captionsetup{justification=raggedright,singlelinecheck=false} 
\begin{minted}[ frame=single, breaklines, breakanywhere, autogobble, fontsize=\small, xleftmargin=0em, framesep=0.2mm, tabsize=2, samepage=true,  highlightcolor=hl, highlightlines={3,25} ]{python}
def specfix(problem, model, C=20, E=10, max_attempts=3):
    rep_model = Persistent(model, "/cache/path")
    ind_model = Independent(rep_model)

    def score(D):
        prompt = CODE_GEN_TMPL.format(description=D)
        # generate C programs, always the same
        # for the same requirements
        progs = islice(rep_model.sample(prompt), C)

        prompt = TEST_GEN_TMPL.format(description=D)
        # generate E tests, always the same
        # for the same requirements
        tests = islice(rep_model.sample(prompt), E)
        
        clusters = cluster(progs, tests)
        return evaluate_entropy_and_consistency(clusters)

    D = problem.description
        
    for attempt in range(max_attempts):
        prompt = REPAIR_TMPL.format(description=D)
        # repairs are always independent, but the sequence
        # of repairs is cached on disk, and replayed
        D2 = next(ind_model.sample(prompt))

        if score(D2) > score(D):
            D = D2
        if score(D) >= THRESHOLD:
            return D
    return D
\end{minted}
\vspace{-1.1\baselineskip}
\end{listing}

To quantify \projName's impact on \textit{SpecFix}, we conducted eight controlled runs on 100 randomly selected HumanEval+ problems using \texttt{gpt-4.1-mini} at temperature~1.0.
Each run varied the caching configuration and the program sample size~($C$) and test sample size~($E$), which estimate behavioral diversity, as summarized in Table~\ref{tab:mnimi_experiment}.
The first run, labeled \textbf{Baseline}, mirrors the original \textit{SpecFix} behavior as no cached responses are used, with the small change of writing all outputs to the cache for future reuse.
\textbf{No-cache} runs, in contrast, ignore caching entirely (no reads or writes), incurring full cost and runtime on every execution and forfeiting reproducibility.

Enabling reading from the baseline’s cached responses immediately reduced query cost to zero on the next run, replaying results byte-for-byte.
When the program sample size increased from $C{=}20$ to $C{=}40$, or from $E{=}10$ to $E{=}20$  the \textbf{First cached} run incurred additional cost only for the 20 new programs and the 10 extra tests respectively. The next \textbf{Cached replays} again dropped to zero cost, demonstrating perfect reuse once the cache was complete.
Without caching, all generations are recomputed from scratch, greatly increasing both cost and runtime.
This confirms \projName’s incremental cost model: identical workloads replay deterministically at zero cost, and only new generations trigger additional queries.
Moreover, reproducible cached runs allowed all experiments to be re-run in seconds without recontacting the API.
Reproducibility is a critical challenge in empirical computer science, especially for tools like SpecFix that rely on stochastic processes such as sampling from LLMs. Without mechanisms like caching and deterministic replay, researchers must rerun entire experiments from scratch, making replication studies prohibitively expensive in time and cost, and exacerbating the replication crisis in the field~\cite{cockburn2020threats}. Local caching addresses this issue by reducing computational overhead, enabling stable and deterministic outcomes, and facilitating data sharing for validation and comparison.
\vspace{-2mm}
\section{Discussion}

While \projName enhances reproducibility and reduces costs at negligible overhead, several challenges remain open.
Future extensions could integrate semantic caching~\cite{meancache2024}, allowing re-use across paraphrased or structurally equivalent prompts while still preserving statistical guarantees.
Introducing concurrency-aware cache access would further accelerate large-scale multi-threaded workflows, enabling safe parallelism across repair attempts and datasets.  



The current implementation has limitations. Its correctness relies on deterministic serialization of prompts and settings; any nondeterminism, such as API-side randomness or floating-point ordering, can introduce subtle drift. Cache storage may grow large for tasks with high prompt diversity, and although disk-based persistence is efficient, long-term cache management and eviction remain manual.

\section{Related Work}

\paragraph{\textbf{Traditional Caching.}} Caching is crucial for system optimization but is challenging to implement correctly. Besnard et al.~\cite{besnard2019memoization} highlighted that caching non-deterministic functions can alter program behavior and proposed memoizing only pure, deterministic computations. This principle extends to web systems, where HTTP caching and CDN policies often exclude dynamic content. Bos et al.~\cite{bos2009cachecard} introduced CacheCard, a NIC-level caching system that avoids caching dynamic web pages unless dependencies are explicitly tracked. In databases, Garrod et al.~\cite{garrod2008ferdinand} described a distributed query result cache that ensures semantic correctness through invalidation triggered by database writes.

\paragraph{\textbf{Probabilistic Sampling and Lazy Evaluation.}} Functional programming research has explored how to represent stochastic computations as reproducible, composable data structures. Scibior et al.~\cite{scibior2015practical} formalize probabilistic sampling as lazy sequences, enabling repeatable access to random draws within purely functional programs.  
This serves as an inspiration for \projName that uses infinite lazy sequences to implement composable layers representing different statistical independence constraints.


\paragraph{\textbf{Frameworks for LLM Programming and LLM Caching.}} Due to existence of LLM-related implementations, LLM-integrated scripts are facing challenges in terms of performance, modularity, and readability. Wang~\cite{LMPL:Wang25} proposed the use of composable effect handling to separate workflow logic from LLM-related effectful operations, thereby enabling modularity without sacrificing the potential for performance optimization. Consistent with this, our work also aims to benefit LLM programming by decoupling LLM caching from core program logic. Frameworks like LangChain~\cite{langchain2023} use string-based caching, where identical prompts return stored outputs to avoid redundant queries, improving efficiency but lacking control over statistical independence. Meanwhile, MeanCache~\cite{meancache2024} and GPTCache~\cite{bang-2023-gptcache} extend caching to semantically similar prompts, complementing \projName. Frameworks like FloTorch~\cite{flotorch2025} optimize agentic workflows by caching intermediate steps with per-step policies. Low-level optimizations, such as vLLM’s key–value cache~\cite{kwon2023vllm}, accelerate token-level computation reuse but do not provide statistical independence control. Cache\ Saver~\cite{potamitiscache} introduces ``namespaces'', where samples labelled with the same namespace are guaranteed to be  independent. Section~\ref{sec:examples} shows examples that are hard to express using namespaces, but are concisely expressible in \projName.

\section{Conclusion}

This work introduces \projName, a cache design pattern that ensures statistical integrity while supporting modular LLM workflows. It encodes statistical independence constraints through reference types: repeatable model references yield identical sequences of responses for a given prompt, while independent references produce fresh, independent sequences of responses. By allowing flexible transformations between reference types, \projName simplifies maintaining statistical integrity across algorithm scopes. A case study on the SpecFix tool demonstrates how \projName reduces research costs while enhancing reproducibility and experimentation.

\bibliographystyle{ACM-Reference-Format}
\bibliography{refs}


\appendix


\end{document}